\documentclass[aps, prd, 10pt, nofootinbib, showpacs, showkeys, superscriptaddress, preprintnumbers, altaffilletter, floatfix, twocolumn, longbibliography]{revtex4-2}

\usepackage[dvipsnames]{xcolor}
\usepackage{xparse}
\usepackage{graphicx}
\usepackage{amsmath,amssymb,amsfonts,amsbsy}
\usepackage{dcolumn}
\usepackage{rotating}
\usepackage{multirow}
\usepackage{makecell}
\usepackage{comment}
\usepackage{overpic}
\usepackage{lipsum}
\usepackage{verbatim}
\usepackage{cancel}
\usepackage{makecell}
\usepackage[english]{babel}
\usepackage[utf8]{inputenc}
\usepackage[normalem]{ulem} 
\usepackage{silence}
\usepackage{url}
\definecolor{plotpurple}{HTML}{5E3C83}
\usepackage[colorlinks=true, allcolors=plotpurple]{hyperref}

\makeatletter
\newcommand*{\rom}[1]{\expandafter\@slowromancap\romannumeral #1@}
\makeatother

\allowdisplaybreaks

\newif\ifshowdraft
\showdrafttrue      

\NewDocumentEnvironment{finaltext}{+b}{%
  \ifshowdraft\else
    {\begingroup\color{black}#1\endgroup}%
  \fi
}{}

\NewDocumentEnvironment{drafttext}{+b}{%
  \ifshowdraft
    {\begingroup\small\color{blue}\begin{quote}\textbf{[Draft note:]}~#1\end{quote}\endgroup}%
  \fi
}{}

\begin{document}

\title{Geometric acceleration in \texorpdfstring{$f(Q,C)$}{f(Q,C)} theories}

\author{Mikel Artola}
\email{mikel.artola@ehu.eus}
\author{Ismael Ayuso}
\email{ismael.ayuso@ehu.eus}
\author{Ruth Lazkoz}
\email{ruth.lazkoz@ehu.eus}
\affiliation{Department of Physics, University of the Basque Country, 48940 Leioa, Spain}
\affiliation{EHU Quantum Center, University of the Basque Country UPV/EHU, 48940 Leioa, Spain}

\author{Gonzalo Olmo}
\email{gonzalo.olmo@ific.uv.es}
\affiliation{Instituto de Física Corpuscular (IFIC), CSIC‐Universitat de Val\`encia, Spain}
\affiliation{Universidade Federal do Ceará (UFC), Departamento de Física, Campus do Pici, Fortaleza - CE, C.P. 6030, 60455-760 - Brazil}

\author{Vincenzo Salzano}
\email{vincenzo.salzano@usz.edu.pl}
\affiliation{Institute of Physics, University of Szczecin, Wielkopolska 15, 70-451 Szczecin, Poland}

\date{\today}

\begin{abstract}
    The $f(Q,C)$ framework of gravity enables the depiction of an effective dark energy fluid that emerges from geometry itself, thus leading to modifications in the cosmological phenomenology of General Relativity. We pursue this approach to discover new and observationally supported (effective) evolving dark energy models. We propose a general $f(Q,C)$ formulation that cannot be simply split into separate functions of $Q$ and $C$, yet it still results in second-order field equations. By employing a particular type of connection, we derive guidelines for new cosmological models, including a variant of the DGP model that appears to be statistically favored over $\Lambda$CDM. Notably, we also demonstrate how to translate solutions within this $f(Q,C)$ framework to $f(Q)$ counterparts at the background level.
\end{abstract}

\maketitle

\section{Introduction}
Gravity, though the weakest of the fundamental interactions, dictates the architecture and fate of the largest cosmic structures. Through its subtle but persistent influence we trace the rhythm of cosmic expansion, yet our understanding of its underlying mechanism remains uncertain and possibly incomplete. Einstein's General Relativity (GR) interweaves the geometry of spacetime with the dynamics of matter and energy, passing all local and strong-field tests with exquisite precision~\cite{Will:2014kxa}. On cosmological scales, however, the theory requires two unseen components to account for structure formation and the observed acceleration of the Universe, presumably cold dark matter (CDM) and dark energy~\cite{Peebles:2002gy, Padmanabhan:2002ji, Weinberg:1988cp}. The concordance model, $\Lambda$CDM, reproduces the data but leaves open the origin of cosmic acceleration and the near-cancellation of vacuum energy. These persistent riddles suggest that the gravitational sector itself might be hiding additional physics.

Those shortcomings have renewed the search for extensions of Einstein's framework~\cite{Clifton:2011jh, Capozziello:2011et, Nojiri:2017ncd}. Early attempts to unify gravity with other interactions through additional dimensions evolved into geometric reformulations that expose new dynamical degrees of freedom within spacetime itself. Among these, the \emph{metric-affine} viewpoint offers a direct generalization: it relieves the Levi-Civita connection from its constraint and treats the affine connection as an independent field possessing its own dynamics. Gravitational information can then be expressed through curvature, torsion, or non-metricity~\cite{BeltranJimenez:2017tkd, BeltranJimenez:2019esp}, three equivalent formulations related by total divergences: $\mathring{R} = -T + B = Q + C$, with $B = 2\mathring{\nabla}_\mu T^{\mu}$ and $C = - \mathring{\nabla}_\mu (Q^{\mu} - \tilde Q^{\mu})$ (quantities with $(\,\mathring{}\,)$ correspond to the symmetric and torsionless framework) contributing only as surface terms in the action. Here $T^{\mu} \!\equiv\! T^{\nu}{}_{\nu}{}^{\mu}$ is the torsion vector, while $Q_{\mu}\!\equiv\!Q_{\mu\nu}{}^{\nu}$ and $\tilde Q_{\mu}\!\equiv\!Q^{\nu}{}_{\mu\nu}$ are the two independent traces of the non-metricity tensor. GR is recovered in any of these representations; differences arise only when non-linear functions of these scalars, such as $f(T)$ or $f(Q,C)$, are considered.

This observation motivates the construction of generalized $f(Q,C)$ theories, in which both the non-metricity scalar $Q$ and its boundary partner $C$ enter the Lagrangian~\cite{Capozziello:2023vne,De:2023xua}. Such models encompass GR and $f(Q)$ gravity as special cases, but display richer dynamics capable of producing self-acceleration (see e.g.~\cite{Pradhan:2023oqo, Paliathanasis:2023kqs, Paliathanasis:2023gfq, Gadbail:2023mvu, Lohakare:2024oeu, Maurya:2024qtu, Myrzakulov:2024pms}) and de~Sitter attractors~\cite{Paliathanasis:2024tep} without an explicit cosmological constant (even perhaps phantom crossings). Thus, these extensions provide a geometric explanation for cosmic acceleration, removing the need for dark energy as an external ingredient.

In this investigation we explore new classes of $f(Q,C)$ models within the \emph{Connection~I} formulation~\cite{De:2023xua}, where the connection is auxiliary and all dynamics are encoded in the functional dependence of $f(Q,C)$. We construct rational and asymptotically bounded deformations of canonical $f(Q,C)$ forms, derive the corresponding Friedmann equations, and identify geometries admitting self-accelerating plateaux and GR-like limits. Our results reveal a family of non-trivial \emph{Friedmannian geometries} in which late-time acceleration arises purely from spacetime non-metricity. Furthermore, we subject the new models to state-of-the-art cosmological precision data to conclude there is an statistical  agreement with the observations and at the same time represent viable alternatives that challenge the $\Lambda$CDM paradigm.

\section{\texorpdfstring{$f(Q,C)$}{f(Q,C)} gravity}
The $f(Q,C)$ extension of $f(Q)$ gravity is defined through the action
\begin{equation}
    S =
    \int \mathrm{d}^4x\, \sqrt{-g}\, \left( \frac{1}{2} f(Q, C) \right) + S_\mathrm{NG},
    \label{eq:action}
\end{equation}
where we have made use of the reduced Planck units $8\pi G = c = 1$; $S_\mathrm{NG}$ denotes the action of the material fields, and $g$ is the determinant of the metric tensor. In a spatially flat Friedmann-Lemaître-Robertson-Walker spacetime, different choices of affine connection lead to three non-equivalent but consistent sets of field equations depending on the specification of the connection. We restrict ourselves to the derivation of the equations for Connection~I in~\cite{De:2023xua}, which are equivalent to those of $f(T, B)$ theories. Specifically, this establishes $Q = -6H^2$ and $C = 6(3H^2+\dot H)$, where $H$ is the Hubble function. As a consequence, the modified Friedmann (and Raychaudhuri) equations take the following compact form,
\begin{align}
    3H^2 &=
    \rho_\mathrm{m} + \rho_\mathrm{r} + \rho_\mathrm{DE}, \label{eq:Friedmann_general} \\[.25em]
    2\dot H + 3H^2 &= -(p_\mathrm{m} +p_\mathrm{r}+ p_\mathrm{DE}), \label{eq:Raychaudhuri_general}
\end{align}
under the assumption that the effective dark energy density and pressure are:
\begin{align}
    \rho_\mathrm{DE} &=
    Q f_Q - 3H \dot{f_C} + \frac{1}{2}(C f_C - f - Q); \label{eq:rho_DE} \\[.25em]
    p_\mathrm{DE} &=
    -\rho_\mathrm{DE} + H(2\dot{f_Q} -  3\dot{f_C}) + \ddot{f_C} - 2\dot{H}(1 - f_Q).
    \label{eq:p_DE}
\end{align}
The subscripts ``m'' and ``r'' stand for matter and radiation, respectively; the dots represent derivatives with respect to cosmic time, which are computed after the partial derivatives $f_X = \partial f/\partial X$ with $X = \{Q, C\}$.

\section{Generic model setup}
We introduce $g(Q)$, an unspecified differentiable function of $Q$, so as to be able to consider the following functional form:
\begin{equation}
    f(Q, C) =
    Q + \alpha C/g(Q).
    \label{eq:f(Q,C)_model}
\end{equation}
Computing the relevant derivatives in Eq.~\eqref{eq:rho_DE} and substituting into Friedmann's equation~\eqref{eq:Friedmann_general}, it can be seen that all $\dot{H}$ terms cancel to give
\begin{equation}
    3H^2 =
    (\rho_\mathrm{m} + \rho_\mathrm{r}) + 108\alpha H^4 \frac{g_Q}{g^2}.
    \label{eq:Friedmann}
\end{equation}
The correction term proportional to $\alpha$ acts as an effective geometric dark energy source. Its relevance must grow as the Universe expands, approaching a quasi-constant $H$, which depends on the asymptotic behavior of $g(Q)$ and its derivative. For viability, this term should be negligible at early times to recover GR, evolve smoothly toward acceleration without
instabilities, and admit a stable de~Sitter attractor ($H \!\to\! H_{\mathrm{dS}}$, $\dot H \!\to\! 0$). A consistent background evolution is therefore the minimal requirement before exploring perturbative stability.

Following~\cite{BeltranJimenez:2019tme}, we showed in~\cite{Ayuso:2021vtj} that, in a pure $f(Q)$ theory within the symmetric teleparallel equivalent of GR framework (STEGR), the left-hand side of the Friedmann equation reduces to a linear first-order inhomogeneous differential equation, which upon the specification
\begin{equation}
    b(Q) =
    \rho_\mathrm{m} + \rho_\mathrm{r},
\end{equation}
can be integrated to give the following structural relation:
\begin{equation}
    f(Q) =
    \sqrt{-Q} \left( \mathcal{M} - \int^{Q} \mathrm{d}x\, \frac{b(x)}{x\sqrt{-x}} \right),
    \label{eq:f(Q)_from_b(Q)}
\end{equation}
with $\mathcal{M}$ an integration constant playing the role of a mass. This means that for a $f(Q,C)$ theory of the sort considered in Eq.~\eqref{eq:f(Q,C)_model}, we can always reinterpret
\begin{equation}
    b(Q) =
    -Q \left(\frac{1}{2} + 3\alpha Q \frac{g_Q}{g^2} \right)
    \label{eq:b(Q)_gen}
\end{equation}
and ascertain the equivalent $f(Q)$ theory, at least in quadratures, by replacing into Eq.~\eqref{eq:f(Q)_from_b(Q)} the expression of $b(Q)$ given by Eq.~\eqref{eq:b(Q)_gen}. Therefore, $f(Q)$-level realizations of the Friedmann equation with interest but discarded on the ground of the ghost problems in $f(Q)$ may acquire renewed viability in a $f(Q,C)$ setup. This can be done by reversing the mapping to give the prescription:
\begin{equation}
    g(Q) =
    6\alpha {\left( \bar{\mathcal{M}} + \int^Q \mathrm{d}x\,\frac{ x + 2b(x) }{x^{2}} \right)}^{-1},
    \label{eq:g(Q)_quadrature}
\end{equation}
with $\bar{\mathcal{M}}$ another constant with mass dimensions.

\section{Preliminary cases}
As different functional forms of $g(Q)$ shape the cosmological dynamics, the next step is to clarify the geometric mechanisms that induce cosmic acceleration in our $f(Q,C)$ framework. To that end, we consider a few representative proposals of limited phenomenological scope. This sets the stage for an intriguing novel model which we test observationally.

\subsection{Simple square root case}
Our first choice is $g(Q) = \sqrt{-Q}$, which in turn gives:
\begin{equation}
    3H^2 =
    (\rho_\mathrm{m} + \rho_\mathrm{r}) - \frac{{3\sqrt{6}\alpha}}{2}H.
    \label{eq:DGP}
\end{equation}
This framework is directly comparable to the Dvali-Gabadadze-Porrati (DGP) braneworld scenario~\cite{Dvali:2000hr, Deffayet:2000uy}, in which a linear term in $H$ signals the transition between a four-dimensional general relativistic regime and an effectively higher-dimensional phase. In this context, the renowned crossover scale is recovered by setting $r_\mathrm{c}^{-1} \!=\! \sqrt{6}\alpha/2$. Nevertheless, in this framework no extra dimensions are introduced: the same pattern emerges purely from the geometric structure encoded in $f(Q,C)$. This behavior illustrates how certain modified frameworks can \emph{mimic geometric leakage effects} without invoking extra dimensions or branes. The appearance of a DGP-like term is a manifestation of the mixed dynamics between non-metricity and boundary effects. Specifically, it originates from the $C f_C$ coupling in the Friedmann equation, arising from the effective dark energy density in Eq.~\eqref{eq:rho_DE}.

Phenomenologically, the linear term in Eq.~\eqref{eq:DGP} generates self-acceleration even in the absence of a cosmological constant, with the crossover scale $r_\mathrm{c}$ governing the transition between the quasi-GR regime ($H \!\gg\! r_\mathrm{c}^{-1}$) 
and a modified phase dominated by the $H/r_\mathrm{c}$ correction.

\subsection{Power-law case}

The square-root form of $g(Q)$ proposed obviously belongs to a wider class of deformations encoded in a power-law function, $f(Q, C) = Q + \alpha C/(-Q)^n$, where now $g(Q) = (-Q)^n$. This is a simple yet physically sound generalization. Repeating the process of derivative computation and substituting in the field equations leads to:
\begin{equation}
    3H^2 \left( 1 + \frac{6\alpha n} {{(6H^2)}^{{n}}} \right) =
    \rho_\mathrm{m} + \rho_\mathrm{r}.
    \label{eq:Friedmann_powerlaw}
\end{equation}
Note that the effective dark energy source scales as $H^{2(1-n)}$. For $n = 0$ it reduces to the linear case, $f(Q, C) = Q + \alpha C$, yielding an Einstein de~Sitter universe which produces no acceleration. Positive values of $n$ enhance late-time effects, while negative $n$ affect the early Universe evolution, allowing $n$ to interpolate between early- and late-time geometric deformations of the standard background.

In the high-curvature regime ($H \!\gg\! H_0$, with $H_0$ denoting the present Hubble parameter), the correction is suppressed for $n > 0$, recovering GR. As $H^2$ decreases, the term grows and can mimic a cosmological constant, so that $n$ controls the transition from a GR-like early phase to geometric late-time acceleration. The case $n = 1/2$ reproduces the DGP cosmology discussed above, whereas $n = 1$ reduces to a $\Lambda$CDM model with effective cosmological constant $\Lambda_\mathrm{eff} = -3\alpha$. In fact, this latter case corresponds to the STEGR framework with $f(Q) = Q + \mathcal{M}\sqrt{-Q} + 6\alpha$, whose cosmological perturbations from an observational perspective have been analyzed in~\cite{Kolhatkar:2025ubm}.

\section{\label{sec:deformed_DGP} DGP-like deformation}
Among the possible choices for $g(Q)$, we notice one that echoes the DGP spirit in an unexpected way. Consider
\begin{equation}
    g(Q) =
    \frac{\sqrt{-Q}}{\sqrt{\lambda - Q}},
    \label{eq:g(Q)_DGP_damped}
\end{equation}
where $\lambda$ sets a characteristic damping scale. This function behaves smoothly for large curvatures, $\vert Q \vert \!\gg\! \lambda$, approaching unity and recovering standard GR. Conversely, the low-curvature regime depends on the sign of $\lambda$; specifically, for positive $\lambda$ it scales as $\sqrt{-Q/\lambda}$ when $\vert Q \vert \to 0$, whereas for $\lambda < 0$ we find a lower bound $\vert Q_{\min} \vert > \vert \lambda \vert$ imposed by the denominator in Eq.~\eqref{eq:g(Q)_DGP_damped}. In what follows we shall see that this bound gets modified when inspecting the low-curvature regime of the cosmological dynamics.

From the general structure of the Friedmann equation~\eqref{eq:Friedmann} in $f(Q,C)$ theories, and using the needed derivatives, we obtain a compact expression governing the background dynamics:
\begin{equation}
    3H^2 =
    (\rho_\mathrm{m} + \rho_\mathrm{r}) - \frac{3\alpha\lambda H}{2\sqrt{H^2 + \lambda/6}}.
    \label{eq:DGP_damped}
\end{equation}
At first sight, this looks strikingly familiar. The extra term recalls the DGP model, but here the ``leakage'' is modulated by a damping factor $(H^2+\lambda/6)^{-1/2}$, which makes the coupling between the four-dimensional world and its teleparallel ``fake bulk'' weaken as the Hubble rate falls. This single factor is crucial: it removes the pathologies of the DGP self-accelerating branch, corresponding to $\alpha \lambda < 0$, and allows the cosmological constant to emerge dynamically. In fact, an exact de~Sitter solution is a right limit of that specific branch, yielding
\begin{equation}
    H_\mathrm{dS}^2 =
    \frac{1}{12} \left( \vert \lambda \vert \sqrt{1 + 36\alpha^2} - \lambda \right).
    \label{eq:H_dS}
\end{equation}
This limit is fully consistent with Eq.~\eqref{eq:DGP_damped}, and no future singularities arise. Nevertheless, since the present matter and radiation energy densities are non-negligible, the Hubble parameter in the de~Sitter phase must be smaller than the present Hubble parameter, $H_\mathrm{dS}^2 < H_0^2$, leading to the following upper bound:
\begin{equation}
    |\lambda| \sqrt{1 + 36\alpha^2} - \lambda < 
    12H_0^2.
    \label{eq:de_Sitter_constraint}
\end{equation}
For $\lambda > 0$, the parameter $\alpha$ can always be tuned to ensure that the inequality~\eqref{eq:de_Sitter_constraint} holds. Conversely, for $\lambda < 0$ the inequality saturates as $\lambda \to - 6H_0^2$, which in turn implies $\alpha \to 0$. Consequently, we adopt $\lambda > -6H_0^2$ as a physical prior on this parameter. In all cases, note that $H_\mathrm{dS}$ receives a correction from the $\alpha$ parameter that cannot be inferred directly from Eq.~\eqref{eq:g(Q)_DGP_damped}.

We now examine the customary asymptotic regimes and asses their compatibility with the $H_\mathrm{dS}$ conclusion.

\subsection{High-curvature regime}

At early times, when $H^2 \!\gg\! \lambda/6$,
the damping term saturates and the modified Friedmann equation becomes
\begin{equation}
    3H^2 \simeq
    (\rho_\mathrm{m} + \rho_\mathrm{r}) - \frac{3\alpha\lambda}{2},
\end{equation}
noticing that the leading correction acts as a cosmological constant $\Lambda_\mathrm{eff} = -3\alpha\lambda/2$. Thus, in the ultraviolet the Universe naturally approaches a $\Lambda$CDM regime. There is no need to impose a separate dark energy component; the high-curvature limit of the theory itself provides one. The subleading term we omitted, proportional to $1/H^2$ no longer corresponds to mimicry of a DGP ``leakage" but rather defines an effective transition scale $r_c^{-1} \simeq 3\alpha\lambda^2/8$
emerging from the coupling between non-metricity and boundary effects. The Universe behaves as a $\Lambda$CDM model with a memory of its ``fake'' higher-dimensional ancestry.

\subsection{Low-curvature regime}

The low-curvature regime of this cosmology is governed by the sign of $\lambda$. Considering first the case where $\lambda \!>\! 0$, as the expansion slows down the limit $H^2 \!\ll\! \lambda/6$ can be explored. In this regime, the damping factor dominates in Eq.~\eqref{eq:g(Q)_DGP_damped} and the Friedmann equation~\eqref{eq:Friedmann} reads
\begin{equation}
    3H^2 \simeq
    (\rho_\mathrm{m} + \rho_\mathrm{r}) - \frac{3\sqrt{6\lambda}\alpha}{2} H.
    \label{eq:lowH}
\end{equation}
Now, the modification is linear in $H$ but its amplitude is finite, controlled by $\sqrt{\lambda}$, with a crossover scale $r_\mathrm{c}^{-1} = \sqrt{6\lambda}\alpha/2$. This distinguishes the model from the original DGP scenario: the ``extra-dimensional leakage'' term no longer grows unboundedly but instead softens as the Universe expands. At late times, the theory approaches a gently self-accelerating regime---a ``damped DGP'' phase---where different effects combine into producing a smooth de~Sitter attractor. A de~Sitter limit $H_\mathrm{dS} = \sqrt{6\lambda}\vert \alpha \vert/2$ then follows from Eq.~\eqref{eq:lowH} when $\rho_\mathrm{m} + \rho_\mathrm{r} \simeq 0$. This is perfectly compatible with Eq.~\eqref{eq:H_dS}, because the only way to get a small Hubble parameter from that expression is precisely to assume $\vert \alpha\vert \!\ll\! 1$, which leads to the same approximate $H_\mathrm{dS}$.

Conversely, the existence of a low-curvature limit for $\lambda < 0$ depends on the magnitude of $\lambda$. If it remains close to zero and $\alpha$ is also small, then from Eq.~\eqref{eq:DGP_damped} it follows that the de~Sitter limit corresponds to a low-curvature regime with $H \!\ll\! H_0$. Nevertheless, as $\lambda \to -6H_0^2$ from above---which, according to Eq.~\eqref{eq:H_dS}, implies $\alpha \to 0$---the asymptotic Hubble parameter remains close to the present value $H_0$, and the aforementioned low-curvature limit can then be reinterpreted as $H \lesssim H_0$.

\subsection{Brief summary}

The physical interpretation of this model is that it behaves as a tempered DGP cosmology, in which the characteristic leakage of gravity into the extra dimension is modulated rather than abrupt. At high-curvature, the model asymptotically mimics the effect of a cosmological constant, producing a geometric correction that is indistinguishable from $\Lambda$CDM at early times. In the low-curvature regime, corresponding to large cosmic scales or late-time dilution of matter and radiation, the extra-dimensional imprint fades smoothly, avoiding the pathologies typically associated with the DGP self-accelerating branch. The parameters $\alpha$ and $\lambda$ respectively control the strength of the gravitational leakage and the scale of its exponential damping, thus governing how and when higher-dimensional effects become relevant. In the limit $\lambda \to \infty$, Eq.~\eqref{eq:DGP_damped} recovers the standard DGP scenario, with an unsuppressed leakage term, whereas a finite $\lambda$ regularises the self-acceleration, ensuring a milder departure from four-dimensional gravity and yielding a stable, and possibly ghost-free, cosmic expansion consistent with current observational bounds.

\section{Observational analysis}
Fixing momentarily $H = H_0$ in Eq.~\eqref{eq:DGP_damped} imposes a normalization condition on the parameters and hints at defining $\Omega_{\lambda}=\lambda/(6H_0^2)$, so that
\begin{equation}
    3\alpha \Omega_{\lambda} =
    \sqrt{1 + \Omega_{\lambda}}(\Omega_\mathrm{m} + \Omega_\mathrm{r} - 1).
    \label{eq:DGP_damped_normalization}
\end{equation}
The matter and radiation fractional densities $\Omega_\mathrm{m}$ and $\Omega_\mathrm{r}$ are expected to remain close to their GR values, so assuming $\Omega_\mathrm{m} + \Omega_\mathrm{r} < 1$ naturally leads to $\alpha \Omega_\lambda<0$, consistent with the accelerating branch aforementioned. Using the constraint~\eqref{eq:DGP_damped_normalization}, we conclude that the Friedmann equation for any redshift reads:
\begin{align}
    E(z)^2 =\,&
    \sqrt{1 + \Omega_{\lambda}} (1 - \Omega_\mathrm{m} - \Omega_\mathrm{r}) \frac{E(z)}{\sqrt{E(z)^2 + \Omega_{\lambda}}} 
    \notag\\[.25em]%
    &+ \Omega_\mathrm{m}(1+z)^3 + \Omega_\mathrm{r}(1+z)^4,
\label{eq:DGPdamped_E(z)}
\end{align}
where we have defined $E(z) \equiv H(z)/H_0$ following the standard notation. The background evolution is modified only by the additional parameter $\Omega_\lambda$, which we restrict to $\Omega_\lambda > -1$. This choice is consistent with the previous discussion of the de~Sitter phase and, moreover, prevents singularities in the Friedmann equation~\eqref{eq:DGPdamped_E(z)}.

Note that both matter and radiation have the usual scaling with the redshift $z \equiv a^{-1} - 1$, as the Friedmann and Raychaudhuri equations~\eqref{eq:Friedmann_general} and~\eqref{eq:Raychaudhuri_general} are perfectly compatible with the customary conservation equation:
\begin{equation}
    \dot{\rho}_i + 3H(\rho_i + p_i) =
    0,
    \label{eq:conservation_equation}
\end{equation}
which guarantees $\rho_\mathrm{m} \propto a^{-3}$ for pressureless matter and $\rho_\mathrm{r} \propto a^{-4}$ for radiation. Hence, any deviation from the standard expansion arises purely from the geometric sector of the theory, rather than from non-standard matter couplings. This prescription allows for a well-defined evolution for the effective dark-energy equation of state, $w_\mathrm{DE} \equiv p_\mathrm{DE}/\rho_\mathrm{DE}$. Both the energy density and the pressure follow from Eqs.~\eqref{eq:rho_DE} and~\eqref{eq:p_DE} under the $f(Q,C)$ prescription of Eq.~\eqref{eq:f(Q,C)_model} and the proposed $g(Q)$ in Eq.~\eqref{eq:g(Q)_DGP_damped}. The explicit formulation of $w_\mathrm{DE}$ is omitted since it provides little analytical insight; nevertheless, we have verified that for $\Omega_\lambda > 0$ it remains above the phantom divide line ($w_\Lambda = -1$), while the opposite occurs for $\Omega_\lambda < 0$.

\subsection{Datasets}

We asses the observational viability of the DGP-like deformation proposed by performing a Bayesian analysis with the most up-to-date datasets. Specifically, we rely on Baryon Acoustic Oscillations (BAO), Cosmic Chronometers (CC), the Cosmic Microwave Background (CMB) and Type Ia Supernovae (SNeIa).

\subsubsection{BAO} 

We employ measurements of the cosmological distances $(D_\mathrm{V}/r_\mathrm{d}, D_\mathrm{M}/r_\mathrm{d}, D_\mathrm{H}/r_\mathrm{d})$ over the range $0.295 \!\leq\! z_\mathrm{eff} \leq 2.3$ from the Data Release 2 of DESI~\cite{DESI:2025zgx}. The sound horizon $r_\mathrm{d} \equiv r_\mathrm{s}(z_\mathrm{d})$ is computed using its integral expression---which remains unmodified as the model has a standard $\Lambda$CDM-like early-time evolution---and implementing the redshift at the drag epoch $z_\mathrm{d}$ provided in~\cite{Aizpuru:2021vhd}.

\subsubsection{CC} 

We include the measurements of $H(z)$ from~\cite{Moresco:2022phi} over the range $0 < z < 1.965$, obtained by using the differential age methods applied to passively evolving galaxies~\cite{Moresco:2022phi, Jimenez:2001gg, Moresco:2010wh, Moresco:2018xdr, Moresco:2020fbm}. As the main source of systematic uncertainties arises from the model assumed to reconstruct these measurements, we also implemented the full covariance matrix following the steps presented in~\cite{Moresco:2020fbm}.

\subsubsection{CMB} 

The compressed CMB likelihood from~\cite{Bansal:2025ipo} is used, imposing a multivariate Gaussian prior over $(R,\,l_a,\,\Omega_\mathrm{b}h^2)$; $R$ and $l_a$ are the shift parameters~\cite{Wang:2007mza}, $\Omega_\mathrm{b}$ is the fractional baryon density and $h = H_0 / (100~ \mathrm{km}\, \mathrm{s}^{-1}\, \mathrm{Mpc}^{-1}$). The use of these priors is justified since the damped DGP model enters in a $\Lambda$CDM regime at early-times.

\subsubsection{SNeIa} 

We consider the Pantheon+ compilation~\cite{Scolnic:2021amr, Peterson:2021hel, Carr:2021lcj, Brout:2022vxf}, selecting SNeIa in the range $0.01 < z < 2.26$. Low redshift events at $z < 0.01$ are not considered in order to minimize the effects of peculiar velocities~\cite{Peterson:2021hel}. In addition, Cepheids calibrated by the \emph{SH0ES}'s team are excluded, as they are deemed incompatible with CMB constraints~\cite{Brout:2022vxf}. Consequently, since $H_0$ and the fiducial absolute magnitude $M$ of SNeIa are degenerate, we analytically marginalize over $M$ following the procedure in~\cite{Conley_2010}.

\subsection{Results}

The sampling is performed over the cosmological parameters $(\Omega_\mathrm{m}, H_0, \Omega_\mathrm{b}, \Omega_\lambda)$ using a Monte Carlo Markov Chain (MCMC) pipeline, assigning to each parameter the following uninformative flat priors based on those employed by the DESI collaboration~\cite{DESI:2025zgx}: $\Omega_\mathrm{m} \in [0.01,0.99]$, $H_0 \in [20,100]$, $\Omega_\mathrm{b} \in [0.005, 0.1]$, and the theoretically motivated $\Omega_\lambda > -1$. This model is then compared with the $\Lambda$CDM model via the Bayes factor $\ln \mathcal{B}_{\Lambda\mathrm{CDM}}^\mathrm{DGPlike}$, computed from the quotient of the evidences using the Nested Sampling algorithm presented in~\cite{Mukherjee:2005wg}. Positive values indicate evidence in favor of the DGP-like deformed cosmology, and the statistical significance is assessed following Jeffreys' criteria~\cite{Jeffreys61}. 

The analysis yields values of the cosmological parameters of the modified gravity proposal consistent with $\Lambda$CDM within $2\sigma$. For this specific model, our pipeline gives the results presented on Table~\ref{tab:results}. The best fit for the DGP-like deformation exhibits some noteworthy features. Compared to $\Lambda$CDM, the analysis yields very similar values of $\Omega_\mathrm{m}$ and $\Omega_\mathrm{b}$, together with a slightly lower Hubble constant $H_0$ and a non-vanishing contribution from the new parameter $\Omega_\lambda$. The corresponding best fit achieves a lower $\chi^2_\mathrm{min}$ than that of $\Lambda$CDM, implying $\Delta\chi^2_\mathrm{min} = -1.04$ in favor of this new model. Bayesian model comparison indicates a mild preference for the deformed scenario, with $\ln \mathcal{B}_{\Lambda\mathrm{CDM}}^\mathrm{DGPlike}$ corresponding to \emph{weak evidence} on Jeffreys' scale~\cite{Jeffreys61}.

\begin{table}[t]
    \centering
    \setlength{\tabcolsep}{.75em}
    \begin{tabular}{ccc}
        \hline\hline\rule{0pt}{1.em}%
        Observable & $\Lambda$CDM & Damped DGP \\\hline\rule{0pt}{1.1em}%
        $\Omega_\mathrm{m}$ & $0.3019_{-0.0035}^{+0.0037}$ & $0.3065_{-0.0052}^{+0.0052}$ \\[.5em]
        $H_0$ & $68.28_{-0.28}^{+0.28}$ & $67.64_{-0.57}^{+0.58}$ \\[.5em]
        $\Omega_\mathrm{b}$ & $0.04836_{-0.00031}^{+0.00032}$ & $0.04936_{-0.00087}^{+0.00087}$ \\[.5em]
        $\Omega_\lambda$ & \textemdash---& $0.20_{-0.17}^{+0.20}$ \\[.5em]
        $\chi_\mathrm{min}^2$ & $1448.07$ & $1447.03$ \\[.5em]
        $\ln \mathcal{B}_{\Lambda\mathrm{CDM}}^\mathrm{DGPlike}$ & \textemdash---& $0.18$ \\[.1em]
        \hline\hline
    \end{tabular}
    \caption{Best fits (specifically the medians) and $1\sigma$ uncertainties of the parameters for the $\Lambda$CDM and the damped DGP cosmology, together with the minimum $\chi^2$ and the logarithm of the Bayes factor.}
    \label{tab:results}
\end{table}

\begin{figure}[t]
    \centering
    \includegraphics{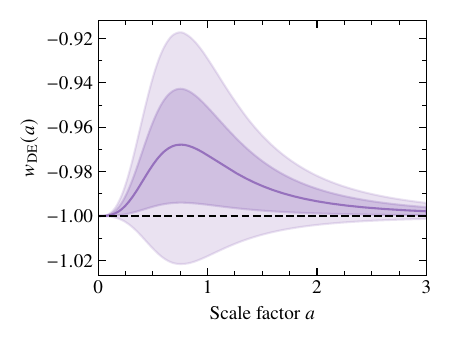}
    \vspace{-1.5em}
    \caption{Effective equation of state parameter associated to the geometric corrections of the damped DGP model. The solid line represents the median at each value of the scale factor, whereas the shaded regions encompass the $68\%$ and $95\%$ confidence regions; the black dashed lines indicate the cosmological constant equation of state, $w_\Lambda = -1$.}
    \label{fig:w_DE}
\end{figure}

In Figure~\ref{fig:w_DE} we show the equation of state parameter of the effective dark energy, $w_\mathrm{DE}(a)$. The plot confirms that $w_\mathrm{DE}$ is fully compatible with $\Lambda$CDM in the high-curvature limit $a \to 0$, as discussed in Sec.~\ref{sec:deformed_DGP}, while the de~Sitter limit can be inferred from the large $a$ behavior. Over all the cosmic history, the median and $1\sigma$ uncertainties lie above the cosmological constant value (black dashed line), reflecting the positive $\Omega_\lambda$ values within the $68\%$ confidence region. Conversely, the $2\sigma$ interval indicates that $w_\mathrm{DE}$ could fall below the phantom divide, allowing for negative $\Omega_\lambda$ although with a lower statistical significance. Interestingly, the equation of state reaches a maximum at $z_{\max} = 0.329_{-0.010}^{+0.015}$, well withing the range of local redshift measurements, where the spread in $w_\mathrm{DE}$ is also the largest. This redshift marks the point where the model departs most from standard $\Lambda$CDM, suggesting that the deviation itself drives the increased uncertainty.

\section{Conclusions}
We have explored a class of $f(Q,C)$ theories that produce late-time cosmic acceleration purely from geometry. Within Connection~I, these models yield second-order field equations and comprise scenarios in which the interplay of non-metricity and boundary terms generates an effective dark energy behavior. A damped DGP-like deformation illustrates how geometric couplings can mimic self-acceleration while avoiding the pathologies of the original model, leading naturally to a smooth de~Sitter attractor. Moreover, we established a background-level correspondence between $f(Q,C)$ and $f(Q)$ frameworks, showing that models discarded in the latter may regain viability in the extended formulation. Confrontation with current cosmological data shows that our proposal is observationally competitive with $\Lambda$CDM. 

Interestingly, we find that the DGP-like deformation leads to a systematically lower value of the present Hubble parameter. This behavior can be understood as a compensation effect: the geometric self-acceleration driven by the mixed $f(Q,C)$ term partially replaces the role of a cosmological constant, thereby enhancing the late-time expansion rate for a given matter density. As a result, the cosmological fit favors smaller values of $H_0$ to reproduce the same distance-redshift relations observed in the data. Our analysis agrees with the observation that lower values of $H_0$ are correlated with larger values of the current equation of state of dark energy ($w_0$) than that of a cosmological constant, $w_0 > -1$~\cite{Colgain:2025nzf}.

Beyond providing a geometrical mechanism for acceleration, $f(Q,C)$ gravity also clarifies the dynamical role of the connection and boundary contributions in symmetric teleparallel frameworks. The appearance of acceleration without exotic fluids highlights the potential of non-metricity as an intrinsic source of cosmic expansion, linked to the geometry rather than to the matter sector. Future work may extend these results by testing other prescriptions along the models here presented to test the robustness of our observational conclusions concerning these models being reasonable contenders for the standard paradigm.

Altogether, our findings support the view that non-metricity and its boundary partner can act as drivers of cosmic acceleration of geometric origin, offering a promising route toward consistent extensions of GR.

\begin{acknowledgments}
MA acknowledges support from the Basque Government Grant No.~PRE\_2024\_1\_0229. MA, IA and RL are supported by the Basque Government Grant IT1628-22, and by Grant PID2021-123226NB-I00 (funded by MCIN/AEI/10.13039/501100011033, by ``ERDF A way of making Europe''). IA is also supported by the Grant Juan de la Cierva funded by MICIU/AEI/10.13039/501100011033 and by “ESF+”. GO has the support from the project i-COOPB23096 (funded by CSIC), and the Grants PID2020-116567GB-C21 and PID2023-149560NB-C21, funded by MCIN/AEI /10.13039/501100011033, and by CEX2023-001292-S funded by MCIU/AEI. 

This article is based upon work from COST Actions CosmoVerse CA21136 and CaLISTA CA21109, supported by COST (European Cooperation in Science and Technology). Portions of the text were generated or refined with the assistance of GPT-5 Plus (OpenAI) for improve the clarity and fluency. All content was subsequently reviewed and verified by the authors.
\end{acknowledgments}

\bibliographystyle{apsrev4-2-titles}
\bibliography{Bibliography}

\end{document}